\documentclass[
reprint,
 amsmath,amssymb,
 aps,
]{revtex4-2}

\usepackage{graphicx}
\usepackage{dcolumn}
\usepackage{bm}

\begin{document}

\preprint{APS/123-QED}

\title{Clocking the Quantum Sojourn Time: Spurious Scatterings and Correction to the Larmor Clock}

\author{Nitish Kumar Gupta}
\email{nitishkg@iitk.ac.in}
\author{A. M. Jayannavar}
\author{S. Anantha Ramakrishna}

\affiliation{Centre for Lasers \& Photonics, Indian Institute of Technology Kanpur, 208016, India}
\affiliation{Institute of Physics, Bhubaneswar, Odisha 751005, India}
\affiliation{Department of Physics, Indian Institute of Technology Kanpur, 208016, India}
\affiliation{CSIR-Central Scientific Instruments Organisation, Chandigarh, 160030, India}

\date{\today}

\begin{abstract}
We revisit the notions of the quantum-mechanical sojourn time in the context of the quantum clocks to enquire whether the sojourn time be clocked without the clock affecting the dynamics of the wave motion. Upon recognizing that the positivity of conditional sojourn time is not ensured even in the case of physically co-evolving clock mechanisms, we trace its origins to the non-trivial inadvertent scattering arising from the disparity, however weak, engendered by the very clock potential. Specifically, our investigations focus on the Larmor spin rotation-based unitary clock where the alleviation of these unphysical contributions has been achieved by correcting the mathematical apparatus of extracting the sojourn times. The corrections have been obtained for both the spin precession-based and spin alignment-based scenarios. The sojourn times so obtained are found to have proper high- and low-energy limits and turn out to be positive definite for an arbitrary potential. The regimen provided here is general and appeals equivalently for unitary as well as non-unitary clocks where the clock-induced perturbations couple to the system Hamiltonian. 
\end{abstract}
\maketitle

\section{Introduction}

One of the longstanding quests in mesoscopic systems is to have a qualitative understanding and a reasonable assessment of time scales involved in a scattering event when a deformable object such as a quantum mechanical wave packet encounters a potential barrier~\cite{landauer1994barrier,hauge1989tunneling,yusofsani2020quantum,sokolovski2018no,muga2007time,brunetti2010time,sokolovski2021tunnelling,rivlin2021determination,dumont2020relativistic,hernandez2021delay,spierings2021tunneling}. Here, interference among the alternative pathways denies objectification of alternatives and a context-free standpoint, stipulating significant efforts even in arriving at a classically veritable definition of time delays. Nonetheless, the dynamical aspects of scattering descriptions continue to attract unabated interest not only on account of prospects of offering alternative viewpoints in quantum mechanics but more so in recent years as it may assist in designing high-speed nanoscale electronic devices; for example, measures such as sojourn time~\cite{jaworski1989concept} can provide us with a time scale to assess whether the dephasing/decoherence would affect the device performance.

The earliest attempt to assess the time-delays in quantum scattering dates to the classic work by Wigner ~\cite{wigner1955lower}. The so-called Eisenbud-Wigner delay time for quasi-monochromatic wave packets is calculated as an energy derivative of the scattering phase shift $\tau_{w}=\hbar{\frac{\partial}{\partial{E}}\theta_{sca.}}$, providing us with an asymptotic time scale for an eventual reflection and transmission. The methodology relies on tracking an identifiable fiducial feature or marker such as the peak of the wave packet as it is scattered off the potential barrier. Evidently, such an approach would be rendered meaningless when scattering brings about strong distortions so much so that the wave packet may lose its very identity. Further, on account of its global property, Eisenbud-Wigner delay time may even churn out negative values, which raise apprehensions about its utility. Indeed, B\"{u}ttiker and Landauer~\cite{buttiker1982traversal} have raised concerns regarding Wigner’s view and emphasized that there exists no causal connection between the incoming and scattered waveforms. The consolidation becomes even more difficult in the case of quantum tunneling as the velocity of propagation itself becomes ill-defined. In another approach, a time of dwell (Smith dwell time) is calculated as the probability of finding the particle inside the spatial region of interest, $\tau_{d}=\int^{\infty}_{-\infty}dt\int^{L}_{0}
 \mid\psi{(x)}\mid^2dx$. This time scale is a statement of the time spent in the region of interest averaged over all the possible pathways, which looks appealing. The problem, however, arises when dealing with evanescent waves as well as in its inability to distinguish between reflection and transmission. Given the inability of various such proposals to conform to our intuitive understanding of traversal timescales, the mathematical artifice of quantum clocks has been invoked, which provides an operational alternative of registering the local time of interaction by attaching an extra degree of freedom (‘clock’) that co-evolves with the sojourning particle. The method relies on external potentials to ensure a meaningful dynamical evolution of the clock when the wave packet is present in the region of interest. The calculation of the expectation value of the associated quantities is then considered to translate into the expectation value of traversal time across the region of space.  The quantum clocks had been proposed as thought experiments, the three prototypical examples of which are - the unitary Larmor clock~\cite{baz1966lifetime,buttiker1983larmor} that depends on the precessional angle accumulated by the spin associated with a particle in a judiciously applied infinitesimal magnetic field; further, we have B\"{u}ttiker-Landauer Oscillating barrier~\cite{buttiker1982traversal} which involves the time-harmonic modulation of the potential barrier and the assessment of traversal time as an adiabatic to non-adiabatic crossover phenomenon; yet another clock, called the non-unitary or e-folding clock~\cite{golub1990modest,buttiker1990traversal,ramakrishna2000imaginary,benjamin2002wave} introduces vanishing imaginary potential over the region of interest to register the logarithmic gradient of the reflection and transmission probabilities. These clocking mechanisms present a direct and physically appealing method for investigating interaction timescales, called the sojourn time. Intuitively we expect the so-clocked sojourn times to be in consonance with our classical perception of time; however, contradictions arise here as well. The emergence of superluminal timescales in barrier tunneling is difficult to explain; even more so are the observations of conditional sojourn times being negative, which raises questions on the expediency of methods and suggests that certain corrections are due to the underlying mathematical machinery. In this context, the recent work by Ramos et al.~\cite{ramos2020measurement}, where they have faithfully implemented the Larmor clock thought experiment, has helped to alleviate concerns regarding the realizability and veracity of clock-based approaches. They have measured the tunneling times for a quantum particle (Bose-Einstein condensate of ultracold rubidium atoms) across a potential barrier and have verified the theoretical predictions of superluminal propagation.
This work is an attempt to make further progress in this direction. We will be working specifically with the periodic Larmor clock to trace out the origins and address the concerns of unphysical characteristics of conditional sojourn times. It turns out that the problem lies in the manner of introduction of clocking mechanism itself; the mismatch and the concomitant scattering at the boundaries of the potential barrier, even in the limits of the vanishing magnetic field, results in spurious perturbing terms which at times engender non-trivial consequences. We then develop a simple and forthright approach to eliminate these unphysical contributions to arrive at the corrected sojourn time.  The treatment presented here is inspired from Ramakrishna et al.~\cite{ramakrishna2002correcting}, who have proposed a novel approach to eliminate spurious scatterings for non-unitary clocks.

\section{Criterion for physically meaningful sojourn time}

In a strict quantum mechanical perspective, the sojourn time is not a dynamical observable of the system, as we lack a qualified Hermitian operator with sojourn time as its eigenvalues. Nonetheless, it is a calculable intermediate quantity with potential utility as an intuitive practical tool in comparing dynamical mesoscopic phenomena, given our classical perceptions about time. The problem in quantum systems, however, arises due to the existence of alternative paths and interference between them, which makes it harder to yield a context-free definition for time or time interval. Hence, before devising a framework for corrected sojourn times, let us first enumerate a compliance criteria on which we will judge the physicality of the outcomes. Since the sojourn time conforms to the description of an ‘interaction time’ so for it to convey any physical meaning, it must not only be real but positive definite as well~\cite{ramakrishna2018long}. Specifically, in the context of physically realizable quantum clocks, a meaningful sojourn time should be :
(i)    Real and positive definite;
(ii)    Additive for non-overlapping regions in space;
(iii)    Calculable and relatable to a measurable quantity;
(iv)    It should possess proper high and low-energy limits (e.g., it should tend to the classically calculable times in high energy limit) 
and (v)    Approaches to B\"{u}ttiker-Landauer time for opaque barriers.
In the low energy scenario, the tunneling time obtained by B\"{u}ttiker and Landauer~\cite{buttiker1982traversal} is very compelling; this timescale sets out a reasonable opaque barrier limit for problems of interest, which any other valid approach should faithfully reproduce. For a particle tunneling through a rectangular potential barrier, the B\"{u}ttiker-Landauer traversal timescale can be defined as 
\begin{equation}
\tau_{BL}=\int^{y_2}_{y_1}\frac{m}{\hbar\kappa(y)}dy=\int^{y_2}_{y_1}\left[\frac{m}{2[V(y)-E]}\right]^{1/2}dy ; 
\end{equation}
where $\hbar\kappa(y)=\sqrt{2m[V(y)-E]}$ represents the instantaneous imaginary momentum of the tunneling particle along Y-axis.

In the coming sections, we provide a prescription for conditional sojourn time that complies with the criteria mentioned above.

\section{System Description: The Larmor Clock}

\begin{figure}[htbp]
\centering
\fbox{\includegraphics[width=\linewidth]{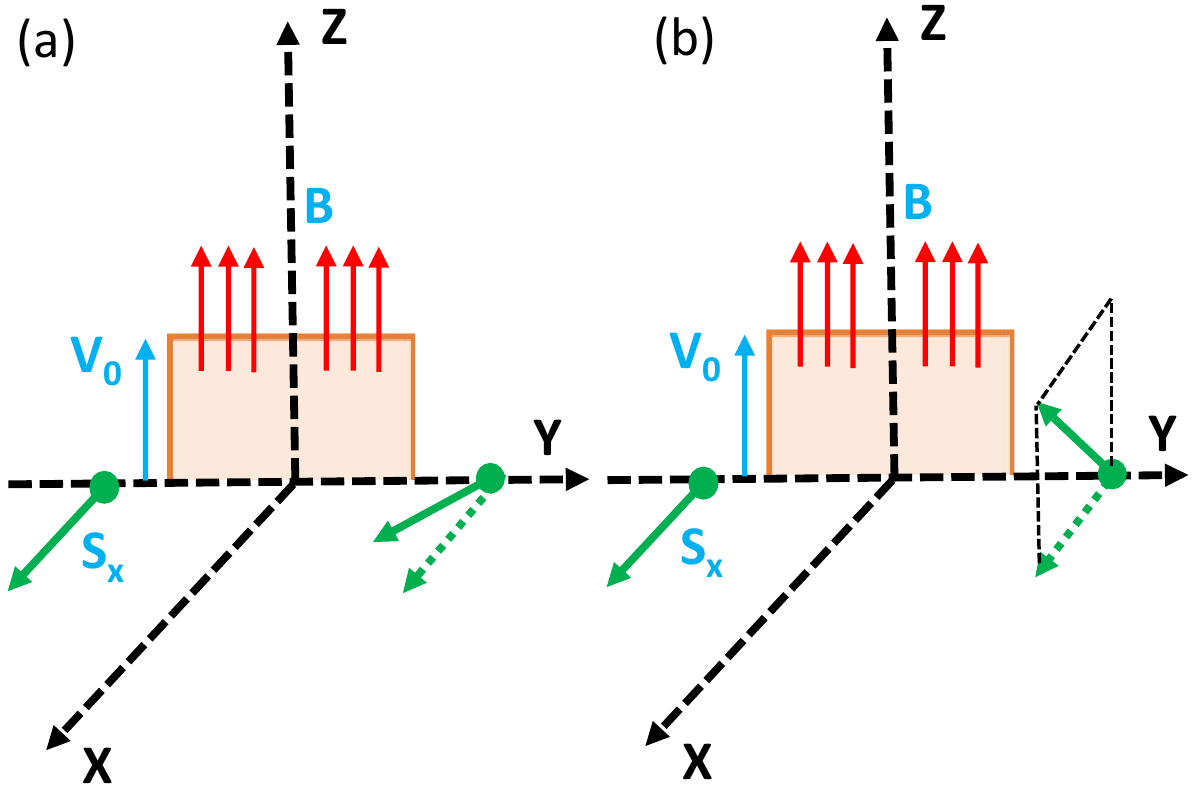}}
\caption{Schematic arrangement of the Larmor clock; spin rotations arising due to spin precession (in XY-plane) and spin alignment (in XZ-plane) have been depicted respectively in (a) and (b) for a particle moving through a potential barrier with local magnetic field.}
\label{fig1}
\end{figure}

Using the concept of Larmor spin precession in an applied magnetic field, Baz and later B\"{u}ttiker had put forward the concept of a unitary clock mechanism that co-evolves with the sojourning particle. From an analysis point of view, the arrangement of the Larmor clock is depicted in Fig.~\ref{fig1}. We consider a spin-$\frac{1}{2}$ quantum mechanical particle to be moving through a potential barrier of height $V(y)=V_0$ and width $l$ along the Y-axis. An extremely weak uniform magnetic field of magnitude $B$ is applied across the potential barrier along the Z-axis, which interacts with the spin degree of freedom of particle to cause traceable changes in the spin state. The particle is initially spin-polarized along X-axis, and as it moves through the magnetic field, the particle experiences spin rotation. This rotation can be attributed to the Larmor precession occurring in the XY-plane (plane perpendicular to the magnetic field), which occurs at a constant Larmor frequency of $\omega_{L}=\frac{g\mu_{B}B}{\hbar}$ (where $\mu_{B}$ is the Bohr magnetron and $g$ is the Land\'{e} $g$-factor ). If the particle takes $\tau_{y}$ time to traverse across the barrier, then the y-spin polarized component in the transmitted flux, approximated to the lowest order in the magnetic field, would be $\langle \boldsymbol{S_{y}} \rangle$ (where $\boldsymbol{\vec{S}}=\frac{\hbar}{2}\boldsymbol{\vec{\sigma}}$ is the spin angular momentum operator). With access to the far side spin, the traversal time (in terms of the spin precession time) is conventionally defined as :

\begin{equation}
\tau_{y}=\frac{2}{g\mu_{B}}\lim_{B \to 0} \frac{\partial \langle \boldsymbol{S_{y}} \rangle}{\partial B} ; 
\end{equation}

B\"{u}ttiker recognized that in addition to the precession, the spin also tends to align with the applied magnetic field $B$ , which causes a rotation parallel to the XZ-plane. This engenders the possibility for another time scale (associated with spin alignment along Z-axis) which is succinctly explored here: In the presence of a localized magnetic field, the interaction Hamiltonian can be written as

\begin{equation}
\boldsymbol{H} = \left\{ \begin{array}{cc} 
               (p^2/2m+V_0)\boldsymbol{I}-(\hbar\omega_{L}/2)\boldsymbol{\sigma_{z}} & \hspace{2mm} \forall~~ -\frac{l}{2}<y<\frac{l}{2} \\
                (p^2/2m)\boldsymbol{I} & \hspace{5mm} elsewhere \\
               
                \end{array} \right.
\end{equation}

where $\boldsymbol{\sigma_{z}}$ is the z-component of Pauli spin matrix. The Hamiltonian $\boldsymbol{H}$ acts upon the spinor $\boldsymbol{\psi}=\begin{pmatrix} \psi_{+}(y)\\ \psi_{+}(y)\end{pmatrix}$, where $\psi_{+}(y)$ and $\psi_{-}(y)$ are the Zeeman components corresponding to parallel and anti-parallel spin states. Since the input spin has been prepared to be spin-polarized along X-axis, it can be written as an equal superposition of these components. Therefore, a projective measurement of $\boldsymbol{\sigma_{z}}$, outside the region of magnetic field, would result $\langle \boldsymbol{S_{z}} \rangle=0$. However, in the presence of a localized magnetic field, the Zeeman energy splitting takes place and the kinetic energy of the two components differs by $\pm\hbar\omega_{L}/2$, leading to a disparity in the wave vectors of the two components. It becomes particularly substantial in case pf barrier penetration where the exponential decay for the two wave functions becomes starkly different:

\begin{equation}
\kappa_{\pm}=\left[\frac{2m}{\hbar^2}(V_0-E)\mp \frac{m\omega_{L}}{\hbar}\right]^{1/2}\approx \left(\kappa\mp\frac{m\omega_{L}}{2\hbar\kappa}\right); 
\end{equation}

where $\kappa=\sqrt{2m[V_0-E]}/\hbar$. Thererfore, it effectuates a mechanism to enhance the tunneling probability of one spin component at the cost of the other and the transmitted flux gets spin-polarized along Z-axis. In the weak magnetic field limit, $\langle \boldsymbol{S_{z}} \rangle$ scales linearly with the magnetic field $B$ , and hence constitutes an equally good candidate for clocking mechanism. Similar to the rationale behind Eq (2), we can define a traversal time corresponding to the spin alignment along the magnetic field as:

\begin{equation}
\tau_{z}=\frac{2}{g\mu_{B}}\lim_{B \to 0} \frac{\partial \langle \boldsymbol{S_{z}} \rangle}{\partial B} ; 
\end{equation}

Thus, the spin evolution through the scattering potential presents two timescales, $\tau_{y}$ and $\tau_{z}$. It turns out that spin precession $\boldsymbol{S_{y}}$ is the dominating mechanism in the case of high-energy propagating particles, while spin rotation $\boldsymbol{S_{z}}$ dominates in low energy barrier penetration.

\section{Problems with the Larmor Clock: Narrative for Corrections}

\begin{figure}[htbp]
\centering
\fbox{\includegraphics[width=\linewidth]{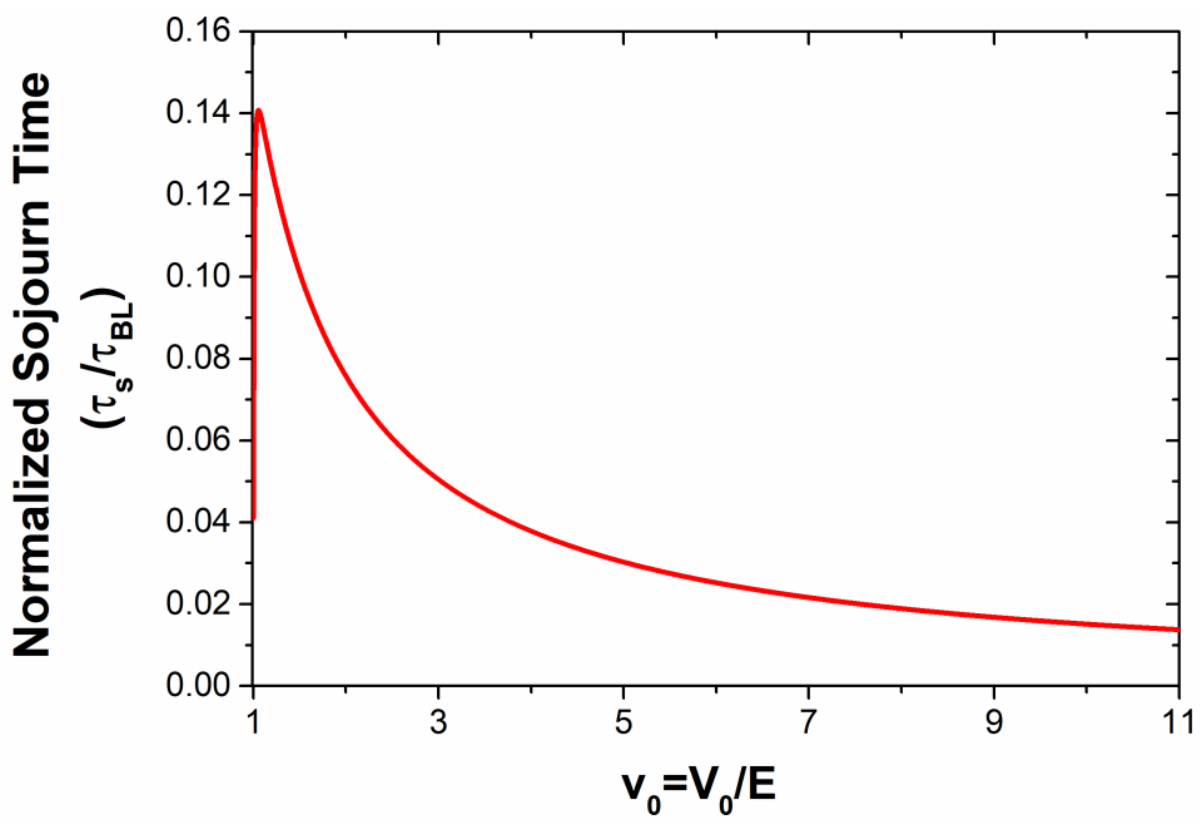}}
\caption{Normalized sojourn time for transmission corresponding to the Larmor clock arrangement of Fig.~\ref{fig1}, in case of barrier penetration.}
\label{fig2}
\end{figure}

Although the concept of registering the traversal times using clocks is physically appealing, at times, it provides results that are difficult to reconcile. Notably, it has been observed that, in general, the clock mechanisms do not yield a positive definite sojourn time. For example, it has been noted that the conditional sojourn time for reflection, for certain deterministic potentials, can come out to be negative~\cite{hauge1989tunneling,golub1990modest}. We have verified that in the case of delta-dimer potential, the reflection sojourn time can become negative.  In the present context, using the definitions provided in the Eq (2) and (5), we have obtained the sojourn time provided by the Larmor clock for the setup of Fig.~\ref{fig1}.  For a judicious comparison, the expression is calculated in the normalized form

\begin{equation}
\frac{\tau_{s}}{\tau_{BL}}=\frac{2(2-v_0)a-v_0/k_0{l}~sin(2ak_0{l})}{4-4v_0+v_0^2sin^2(ak_0{l})}
\end{equation} where $a=\sqrt{1-v_0}$; $v_0=V_0/E$; $\tau_{BL}= ml/{\hbar}k_0\sqrt{|v-1|}$ and $k_0=\sqrt{2mE}/\hbar$.
We observe that the calculated sojourn time refuses to comply with the B\"{u}ttiker-Landauer time and tends to zero in the opaque barrier limit; in addition, the expression for the sojourn time does not ensure a positive definite result.  As earlier pointed out, this violates our criteria for a meaningful sojourn time. To further elaborate, the normalized traversal time has been plotted in Fig.~\ref{fig2} for the case of barrier penetration in low-energy scenarios, where we observe an asymptotically vanishing behavior. Thus, the present machinery for sojourn time calculation provides improper low-energy limits and may bring about unphysical negative outcomes, suggesting that corrections are due in the underneath mathematical apparatus.

\section{Spurious scatterings and their elimination: Larmor Clock Correction}

Our objective here is first to develop an understanding of the origins of spurious scatterings and then invent a strategy to surmount these problems. As we will formally establish, the origins lie in spurious scatterings concomitant with the clock potential, introduced for controllable spin evolution. However, in addition to invoking the desired effects of spin rotation, it alters the scattering dynamics itself in a non-trivial manner. This can be established as follows: As mentioned before, the traversal time can be assessed using spin precession ($\tau_{y}=\frac{2}{g\mu_{B}}\lim_{B \to 0} \frac{\partial \langle \boldsymbol{S_{y}} \rangle}{\partial B}$). The expectation value of y-spin polarized component in the transmitted flux can be found out as

\begin{equation}
\langle \boldsymbol{S_{y}} \rangle=\left\langle \boldsymbol{T} \middle|\boldsymbol{S_{y}} \middle| \boldsymbol{T}\right\rangle=\frac{\hbar}{2}\left\langle \boldsymbol{T} \middle|\boldsymbol{\sigma_{y}} \middle| \boldsymbol{T}\right\rangle;
\end{equation}

 $\boldsymbol{T}$ is the spinor for the transmission
\begin{equation}
\boldsymbol{T}=\frac{1}{\sqrt{|{T_{+}}|^2+|{T_{-}}|^2}}\begin{pmatrix} {T_{+}}\\ {T_{-}}\end{pmatrix}
\end{equation}

where 

\begin{equation}
T_{\pm}=\frac{t_{12}^{\pm}t_{23}^{\pm}e^{ik_{\pm}l}}{1-r_{23}^{\pm}r_{21}^{\pm}e^{2ik_{\pm}l}}
\end{equation}
 
are the overall transmission coefficients.

The partial transmission and reflection coefficients are 

\begin{equation}
t_{12}^{\pm}=\frac{2k_{\pm}}{k_{\pm}+k_{0}};~ t_{23}^{\pm}= \frac{2k_{0}}{k_{\pm}+k_{0}} 
\end{equation}
\begin{equation}
r_{21}^{\pm}=\frac{k_{\pm}-k_{0}}{k_{\pm}+k_{0}};~r_{23}^{\pm}=\frac{k_{\pm}-k_{0}}{k_{\pm}+k_{0}};
\end{equation}

For a rectangular potential well $k=\sqrt{\frac{2m(E-V_0)}{\hbar^2}}$,  $k_{0}=\sqrt{2mE/\hbar^2}$ and $k_{\pm}=k{\mp}\frac{m\omega_{L}}{\hbar{k}}$. 
Now, Eq (7) can be simplified to

\begin{equation}
\begin{aligned}
     \langle \boldsymbol{S_{y}} \rangle & 
     &=-\frac{i\hbar}{2}\frac{T_{+}^{*}T_{-}-T_{-}^{*}T_{+}}{{|{T_{+}}|^2+|{T_{-}}|^2}}=\hbar\frac{{Im}(T_{+}^{*}T_{-})}{{|{T_{+}}|^2+|{T_{-}}|^2}}
\end{aligned}
\end{equation}

Evidently, we observe that in addition to an intended dependence of $\langle \boldsymbol{S_{y}} \rangle$ on the externally applied magnetic field, some inadvertent fixation of $\langle \boldsymbol{S_{y}} \rangle$ also arises (through partial scattering coefficients $r_{ij}$ and $t_{ij}$). In physical settings, such an artifact is avoided by making the applied magnetic field vanishingly small, which is corresponding to the infinitesimal limit in $B$. Naively, here as well, we expect it to ensure that the terms containing $B$ would vanish and be prevented from making any involuntary scattering changes. However, the presence of derivatives in our sojourn time expressions may still permit the first-order terms in $B$ to reflect in the final expressions.  Thus, there persists a subtle problem that some spurious and unphysical scattering contributions also end up reflecting in the calculated values of $\tau_{y}$ and $\tau_{z}$, demanding a course correction. 

By noting down the forms and mannerism of partial wave superposition, we recognize that clocking potential invariably modifies the scattering phenomenon whenever it promulgates an abrupt change in the spatial profile. This suggests the clock potential must be either judiciously apodized, or we need to devise a mathematical workaround to let the clock potential only result in spin rotation without causing inadvertent scatterings. Here, the perturbative structure of the scattering phenomenon convinces us that the magnetic field-induced spin rotation would entail a paired combination $B*l$. The spurious scattering terms (arising from $r_{ij}$ and $t_{ij}$), on the other hand, would rely on unpaired $B$. We utilize this crucial insight for isolating and negating the effects of spurious scattering. We treat $B$ and $B*l= \zeta$ as two independent variables, keeping the $\zeta$ constant we enforce ${B \to 0}$ in the Eq (2). This way, we define a corrected sojourn time for transmission as:

\begin{equation}
\tau_{y}=\frac{2l}{g\mu_{B}}\lim_{\zeta \to 0} \frac{\partial \langle \boldsymbol{S_{y}}(\zeta, B=0)\rangle} {\partial \zeta} ; 
\end{equation}

For this substitution, we also note that as ${B \to 0}$
\begin{equation}
\begin{aligned}
     \exp(ik_{\pm}l) \to \exp(ikl)~\exp\left[\mp{i}\left(\frac{mg\mu_{B}}{\hbar^2{k}}\right)(Bl)\right]  \\
     \\
     =\exp(ikl)~\exp(\mp{i}\alpha\zeta) 
                                    \end{aligned}
\end{equation} 

here $\alpha=\frac{mg\mu_{B}}{\hbar^2{k}}$. With this framework of corrected sojourn time, we proceed to derive the sojourn time expression for the setup depicted in Fig.~\ref{fig1}. To this objective, we first obtain the expression for $\langle \boldsymbol{S_{y}}(\zeta, B=0)\rangle$ using Eq (12).
The numerator of it under prescribed conditions yields:

\begin{equation}
    {Im(T_{+}^{*}T_{-})}=\frac{1}{D}|t_{12}t_{23}|^2~\sin(2\alpha\zeta)(1-|r_{23}r_{21}|^2)
\end{equation}
with $D=|1+|r_{23}r_{21}|^{2}e^{4i\alpha\zeta}- 2Re(r_{23}r_{21}e^{2ikl})e^{2i\alpha\zeta}|^2$.
The denominator also transforms to:

\begin{equation}
\begin{aligned}
 |{T_{+}}|^2+|{T_{-}}|^2 =|t_{12}t_{23}|^2~&\left[\frac{1}{1-r_{23}r_{21}e^{2ikl}e^{2i\alpha\zeta}} \right. \\ &\left.+\frac{1}{1-r_{23}r_{21}e^{2ikl}e^{-2i\alpha\zeta}}\right]
  \end{aligned}
\end{equation}

therefore, by enforcing $B=0$ and retaining the paired combination $B*l= \zeta$, the Eq (12) becomes

\begin{equation}
\begin{aligned}
 \langle \boldsymbol{S_{y}}(\zeta, B=0)\rangle=\\
 \hbar|t_{12}t_{23}|^2(1-|r_{23}r_{21}|^2) &\left[\frac{2~\alpha~{cos(2\alpha\zeta)}}{D(|{T_{+}}|^2+|{T_{-}}|^2)}\right.\\
&\left.+\sin(2\alpha\zeta)\frac{\partial}{\partial \zeta}\left(\frac{1}{D(|{T_{+}}|^2+|{T_{-}}|^2)}  \right) \right]
  \end{aligned}
\end{equation}

Now using Eq (13), we obtain a general expression for the corrected sojourn time based on spin precession 

\begin{equation}
    \tau_{y}=\frac{2\alpha\hbar{l}}{g\mu_{B}}\frac{(1-|r_{23}r_{21}|^2)}{(1+|r_{23}r_{21}|^2)-2Re(r_{23}r_{21}e^{2ikl})}
\end{equation}

Since for any real potential $r_{ij}<1$, therefore the above expression ensures that we always get a positive definite result for transmission sojourn time. Further simplifications corresponding to the arrangement of Fig.~\ref{fig1} provide us with an explicit form of the normalized sojourn time

\begin{equation}
\frac{\tau_{s}}{\tau_{BL}}=\frac{2(2-v_0)a}{4-4v_0+v_0^2sin^2(ak_0{l})}
\end{equation} 
with the nomenclature as employed in Eq (6).

In a similar manner, we define the corrected sojourn time corresponding to the spin alignment as

\begin{equation}
\tau_{z}=\frac{2l}{g\mu_{B}}\lim_{\zeta \to 0} \frac{\partial \langle \boldsymbol{S_{z}}(\zeta, B=0)\rangle} {\partial \zeta} ; 
\end{equation}

Here 

\begin{equation}
\langle \boldsymbol{S_{z}} \rangle=\left\langle \boldsymbol{T} \middle|\boldsymbol{S_{z}} \middle| \boldsymbol{T}\right\rangle=\frac{\hbar}{2}\left\langle \boldsymbol{T} \middle|\boldsymbol{\sigma_{z}} \middle| \boldsymbol{T}\right\rangle;
\end{equation}

or
\begin{equation}
\langle \boldsymbol{S_{z}}\rangle=
\frac{|{T_{+}}|^2-|{T_{-}}|^2}{{|{T_{+}}|^2+|{T_{-}}|^2}};
\end{equation}

\begin{figure}[htbp]
\centering
\fbox{\includegraphics[width=\linewidth]{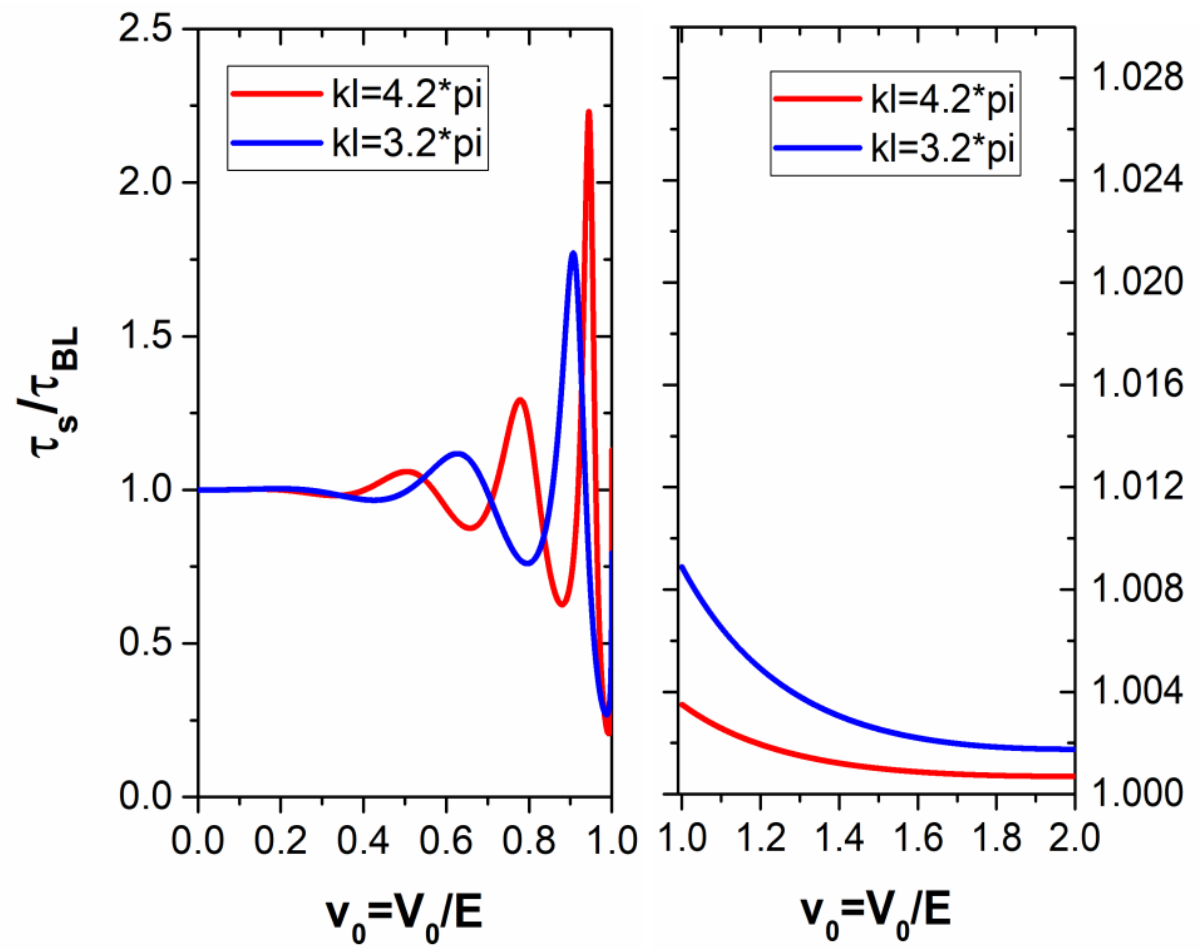}}
\caption{Normalized corrected sojourn time for transmission for the Larmor clock setup of Fig.~\ref{fig1}. The left plot is for above the barrier propagation and the right one is for sub-barrier tunneling.}
\label{fig3}
\end{figure}

Let us define $w=|{T_{+}}|^2+|{T_{-}}|^2$, $u=t_{12}t_{23}e^{-\kappa{l}}$ and $v=r_{23}r_{21}e^{-2\kappa{l}}$, with $k=i\kappa$.

therefore
\begin{equation}
    \begin{aligned}
   \langle \boldsymbol{S_{z}}(\zeta, B=0)\rangle \\
    =\frac{\hbar{|u|^2}}{w~D'}[2(1-|v|^2)~\sinh(2\alpha\zeta)-4Re(v)],
    \end{aligned}
\end{equation}

where $D'=(1+|v|^2{e^{4\alpha\zeta}}-2Re(v)e^{2\alpha\zeta})(1+|v|^2{e^{-4\alpha\zeta}}-2Re(v)e^{-2\alpha\zeta})$. Which leads to the general expression for the corrected sojourn time to be 

\begin{equation}
    \tau_{z}=\frac{ml}{\hbar\kappa}\frac{(1-|r_{23}r_{21}|^2{e^{-4\kappa{l}}})}{(1+|r_{23}r_{21}|^2{e^{-4\kappa{l}}})-2Re(r_{23}r_{21}e^{-2\kappa{l}})}
\end{equation}

For the case of symmetric barrier potential of Fig.~\ref{fig1}, we obtain

\begin{equation}
    \frac{\tau_{s}}{\tau_{BL}}=\frac{(1-e^{-4\kappa{l}})}{(1+{e^{-4\kappa{l}}})-2(1-8(v-1))e^{-2\kappa{l}})}
\end{equation}

The results obtained for the normalized corrected sojourn times have been plotted in Fig.~\ref{fig3} (using Eq (19) and Eq (25)). The normalization has been performed with respect to the B\"{u}ttiker -Landauer traversal time ($\tau_{BL}$) for $E<V_0$ and with respect to the classical Eisenbud-Wigner time for $E>V_0$.

From these results, we notice that the corrected sojourn times are positive definite and also comply with B\"{u}ttiker -Landauer time in the opaque barrier limit. In the high energy limit, on the other hand, the spin precession time ($\tau_y$) approaches the classical Eisenbud-Wigner delay time. Furthermore, it is important to note that these spin rotation-based corrected sojourn times turn out to be identical to the corrected sojourn time obtained for the non-unitary (imaginary) potential method~\cite{ramakrishna2002correcting}. This equivalence is compatible with B\"{u}ttiker's interpretation that spin precession $\boldsymbol{S_{y}}$ is the dominating mechanism in case of wave propagation, while spin rotation $\boldsymbol{S_{z}}$ dominates for wave tunneling.

\section{Conclusion}
To summarize, we have revisited the methodology of sojourn time extraction using the periodic Larmor clock and highlighted the possibility of negative and unphysical outcomes associated with it. It has been established that the problem lies in the mathematical machinery where the very clocking mechanism perturbs the scattering phenomena, originating out of the abrupt potential steps. These contributions do not subside even in vanishing field limit and must be removed to ensure physically meaningful solutions. To abolish these concerns, we have provided a course correction in the mathematical apparatus and devised a solution where the intended spin evolution can be decoupled from the spurious scatterings effects and hence, can be reliably extracted. The corrected sojourn times satisfy the laid out criteria of reasonability and exhibit proper high and low-energy limits. Given recent implementations of the Larmor clock, we believe that our work further supports the efforts of associating a physically meaningful time scale with dynamical mesoscopic phenomena.

\section{Acknowledgement}

A. M. Jayannavar acknowledges DST, India for J C Bose fellowship.

\bibliography{references_aps}

\end{document}